\documentclass[prd,nofootinbib,showpacs,floatfix,
superscriptaddress]{revtex4}
\usepackage{amssymb}
\usepackage{graphics}
\usepackage{amsmath}
\usepackage{dsfont}
\usepackage{amsfonts}
\usepackage{mathrsfs}
\usepackage{bm}
\usepackage[]{latexsym}

\usepackage{graphicx}
\usepackage{epsfig}
\usepackage{bm}
\usepackage[T1]{fontenc}
\usepackage[latin9]{inputenc}
\usepackage{amssymb}
\usepackage{float}
\usepackage{amsmath}
\usepackage{dcolumn}
\usepackage{cancel}
\usepackage[colorlinks]{hyperref}
\usepackage[usenames,dvipsnames]{color}
\hypersetup{
     breaklinks=true,
    pdfstartview={FitH},    
    colorlinks=true,       
    linkcolor=blue,          
    citecolor=red,        
    filecolor=magenta,      
    urlcolor=blue,           
    anchorcolor=green,      
    linktocpage=true
}

\newcommand{\be}{\begin{equation}}\newcommand{\ee}{\end{equation}}
\newcommand{\bea}{\begin{eqnarray}}\newcommand{\eea}{\end{eqnarray}}
\newcommand{\brr}{\begin{array}}\newcommand{\err}{\end{array}}
\newcommand{\bit}{\begin{itemize}}\newcommand{\eit}{\end{itemize}}
\newcommand{\ben}{\begin{enumerate}}\newcommand{\een}{\end{enumerate}}

\newcommand{\ba}{\begin{array}}
\newcommand{\ea}{\end{array}}
\def\lf{\left}

\def\non{\nonumber}

\def\ri{\right}
\def\bp{{\bf {p}}}\def\bk{{\bf {k}}}\def\bx{{\bf {x}}}
\def\by{{\bf {y}}}\def\bq{{\bf {q}}}

\def\lnu{\nu}

\begin{document}
\title{On the flavor/mass dichotomy for mixed neutrinos: a phenomenologically motivated analysis based on lepton charge conservation in neutron decay}

\author{Giuseppe Gaetano Luciano\footnote{email: giuseppegaetano.luciano@udl.cat}} 
\affiliation
{Applied Physics Section of Environmental Science Department, Universitat de Lleida, Av. Jaume
II, 69, 25001 Lleida, Spain}
\date{\today}

\begin{abstract}
\emph{Flavor/mass dichotomy} for mixed fields 
is one of the most controversial issues in Quantum Field Theory. 
In this work we approach the problem by considering mixing
of neutrinos and computing the transition 
amplitude for the paradigmatic neutron $\beta$-decay $n\,\rightarrow\, p \,+\,e^-\,+\,\bar\nu$. Calculations are developed
by utilizing the following different representations 
of neutrino states: \emph{i}) 
Pontecorvo states, \emph{ii}) mass states and \emph{iii})
exact QFT flavor states, which are defined as eigenstates of the flavor charge. 
As a guiding principle, we invoke the conservation
of lepton charge in the interaction vertex, which
is fundamentally inbuilt into the Standard Model at tree level. 
In the short-time limit, we show that the only 
fully consistent scenario is that based on QFT states, whereas 
the other two frameworks contradict the underlying theory.
This result provides a crucial step
towards understanding the r\^ole of neutrino mixing in weak decays
and contributes to elucidate the flavor/mass controversy in QFT. 
\end{abstract}


\date{\today}

\maketitle

\section{Introduction}
Neutrino mixing and oscillations are among the most fascinating, but least understood phenomena in particle physics. 
Although the idea of neutrino oscillations was first proposed 
by Pontecorvo in 1957~\cite{Ponte57}
by analogy with kaons, the theoretical framework in its modern form only dates back to the 1970's~\cite{Grib,Frit}, after the discovery of the second~\cite{Maki} and third~\cite{Third} generation of neutrinos and the later extension to include environmental effects~\cite{Wolf,Mikh}. 
On the experimental side, significant progress was achieved 
between 1998 and 2002 with the detection of atmospheric and solar
neutrino oscillations by Super-Kamiokande~\cite{SuperK,NP} and SNO~\cite{SNO,McD}, shortly thereafter confirmed by KamLand~\cite{KamL}. 

Besides their intrinsic relevance, neutrino oscillations provide to date the only evidence of physics beyond the Standard Model (SM), thus motivating ever-new research both at theoretical and experimental level. In this direction, special focus has been recently placed
on the study of gravitational/acceleration effects on flavor oscillations~\cite{Stodo,Burgard,Fornengo,Cardall,Lambiase,GiuntiKim,Mavro,Paglia,Vissani,Capozziello,Dvo,ETG,BlasSma,Luc,Luc2,Swami,Khalifeh}, with non-trivial results in the framework of extended theories~\cite{Capozziello,ETG}. In parallel, interesting applications have been found in quantum information~\cite{BlasEnt,Alok,Kumar,Roggero,Li,Matrella}, quantum optics~\cite{Marini} and geophysics~\cite{Geo}, where neutrinos are used as a probe to determine the chemical composition of the inner Earth. 

Despite the intensive study in quantum mechanics (QM), comparably 
less attention has been devoted to the analysis of neutrino mixing in Quantum Field Theory (QFT). However, it is the field theoretical investigation that is more pertinent to high-energy phenomena in particle physics, due to the SM being built upon QFT.  
The first consistent treatment of flavor mixing based on the 
construction of Fock spaces for quantized neutrino fields
was developed in~\cite{BV95}, highlighting the shortcomings of the original QM theory. Indeed, 
while Pontecorvo mixing transformations act as pure rotations 
between single-particle states of neutrinos with definite masses (henceforth ``mass states'') and neutrinos with definite flavors (``flavor states''),  their QFT counterparts exhibit the structure of a rotation nested into a non-commuting Bogoliubov transformation~\cite{BV95,Garg}. This leads to orthogonal vacuum states for mass and flavor fields~\cite{BV95}, the latter behaving as a condensate of massive particle/antiparticle pairs with a complex entangled structure~\cite{LucPetr,Cabo}. 
In turn, the Fock spaces built upon these two vacua are \emph{unitarily} (i.e. physically) \emph{inequivalent} to each other, giving rise to characteristic effects that lie outside the domain of QM~\cite{Bare,CapolDark,MavDark,Blasone:2017nbf,Jizba,Quaranta,LucianoTs} (see~\cite{SmVit} for a recent review). We emphasize that similar considerations
had been previously outlined in~\cite{Berna,Berna2} in relativistic QM. 
On mathematical grounds, the theoretical understanding of QFT mixing 
has been confirmed by the rigorous analysis of~\cite{Hannab}. 

In recent years renewed interest in the field theoretical analysis of mixing has been prompted by the study of weak interactions involving 
neutrinos. In particular, a fruitful arena where to test
the QFT formalism was offered by
the weak decay of accelerated protons~\cite{Ahluw,ProtBlas,ProtMat}, 
whose occurrence is allowed by the Unruh effect.
In this context, by invoking consistency with the general covariance of QFT, the scalar decay rate for the proton was computed both in the laboratory frame and in the comoving rest-frame,
with conflicting results on whether to utilize flavor~\cite{Ahluw,ProtBlas} or mass states~\cite{ProtMat} for mixed neutrinos (\emph{flavor/mass dichotomy}, see also~\cite{GiuntiLee,GiuntiUn,Akh,Remarks,SmaldTur,GaetanoLucia} for further discussion on the topic). It must be emphasized that such a controversy is peculiar to QFT~\cite{QFTCon}, whereas in QM the
formal equivalence between the two representations  is 
warranted by the Stone-von Neumann uniqueness theorem~\cite{Stone,VN}. While providing valuable insights on the r\^ole of QFT mixing in weak interactions, these studies
have posed the problem of the very nature of asymptotic neutrinos, 
requiring fresh efforts to sort things out.
In passing, we mention that stimulating analysis of flavor mixing and weak decays were conducted in~\cite{CapW} and~\cite{CYL}. 
 
Starting from the above premises, in this work we
explore more in-depth the inherent features of neutrino mixing
in QFT. Specifically, we attempt to elucidate the aforementioned flavor/mass controversy by focusing on the computation of the transition amplitude for the neutron $\beta$-decay. We
carry out calculations by resorting to three different 
representations of neutrino states: 
\emph{i}) Pontecorvo states, \emph{ii}) mass states and
\emph{iii}) exact QFT flavor states, which are defined as eigenstates of the flavor charge operators. As a result, we obtain
three different expressions for the neutron decay amplitude, 
which only overlap in the limit of ultra-relativistic neutrinos.
The question arises as to which of these descriptions
provides the physically correct scenario. Waiting for experimental
hints that could point us toward the solution, 
here we follow a phenomenologically motivated approach: by requiring the conservation of lepton charge in the interaction vertex, which is empirically observed and built into the SM at tree level~\cite{Cli}, we show that the only fully consistent framework is that founded on QFT states. On the other hand, both Pontecorvo and mass representations 
lead to a violation of lepton charge,  
contradicting the underlying theory. We remark that this conclusion is simply reached on the basis of consistency with SM expectations and does not assume any detector-dependent model for
neutrino oscillations. Such a result should help clarifying the
true nature of asymptotic neutrinos against the variety of exotic claims recently appeared in the literature. 

The remainder of the work is organized as follows: in the next Section we
review the QFT formalism of neutrino mixing. 
Section~\ref{Wip} is devoted to the calculation of the tree-level $\beta$-decay amplitude in the Pontecorvo and mass representations.  Consistency with the conservation of lepton charge is checked by considering the ``short-time limit'' of the above amplitudes, i.e. the limit for small 
distances from the production vertex. The same analysis is then developed with exact QFT flavor states in Sec.~\ref{ExQFTs}. To avoid unnecessary technicalities 
and make the physical insight of our analysis as transparent as possible, 
we neglect flavor changing loop induced processes, which however would not be relevant for our discussion. Conclusions and outlook are summarized in Sec.~\ref{Disc}.
Throughout the whole manuscript, we use natural units $\hbar=c=1$. 

\maketitle

\setcounter{equation}{0}

\section{Neutrino mixing in QFT: mathematical aspects and physical implications}
\label{QFTFo}
Let us start by reviewing the QFT formalism of neutrino mixing and 
the main differences with the QM Pontecorvo treatment (for more details, see~\cite{BV95}). Toward this end, we remind that Pontecorvo mixing transformations for single-particle flavor states read\footnote{In this work we consider a simplified model involving only two flavors. However, the validity of our result is not spoilt by the generalization to the third generation.}
\begin{subequations}
\label{Pontec}
\begin{align}
\label{Ponteca}
|\nu^r_{\bk,e}\rangle_P&=\cos\theta\,|\nu^r_{\bk,1}\rangle+\sin\theta\,|\nu^r_{\bk,2}\rangle\,,\\[2mm]
\label{Pontecb}
|\nu^r_{\bk,\mu}\rangle_P&=-\sin\theta\,|\nu^r_{\bk,1}\rangle+\cos\theta\,|\nu^r_{\bk,2}\rangle\,,
\end{align}
\end{subequations}
where $|\nu^r_{\bk,i}\rangle$, $i=1,2$, are the states of neutrino with definite masses $m_i$, while $|\nu^r_{\bk,\ell}\rangle_P$, $\ell=e,\mu$, are the states of neutrino with definite flavors $\ell$ (electron and muon, respectively). We are assuming equal momentum $\bk$ and helicity $r=1,2$ for neutrinos with different masses. The subscript $P$ is a reminder for Pontecorvo states. 

At level of QFT, the above relations 
are rewritten in the form
\begin{subequations}
\label{Pontecf}
\begin{align}
\label{Pontecfa}
\nu_e(x)&=\cos\theta\,\nu_1(x)+\sin\theta\,\nu_2(x)\,,\\[2mm]
\label{Pontecfb}
\nu_\mu(x)&=-\sin\theta\,\nu_1(x)+\cos\theta\,\nu_2(x)\,,
\end{align}
\end{subequations}
where $\nu_\ell(x)$, $\ell=e,\mu$, are the (interacting) Dirac neutrino fields with definite flavors, while $\nu_i(x)$, $i=1,2$, are the (free) fields with definite masses. These are expanded according to the usual Fourier-decomposition~\cite{BV95}
\be
\label{fe}
\nu_i(x)\,=\,\frac{1}{\sqrt{V}}\sum_{\textbf{k},r}\left[u_{\textbf{k},i}^r\,\alpha^r_{\textbf{k},i}(x^0)\,+\,v_{-\textbf{k},i}^r\,\beta^{r\dagger}_{-\textbf{k},i}(x^0)
\right]e^{i\textbf{k}\cdot\textbf{x}}\,,\qquad i=1,2\,,
\ee
where $\alpha^r_{\textbf{k},i}(x^0)=\alpha^r_{\textbf{k},i}\,e^{-i\omega_{\bk,i}x^0}$, $\beta^{r\dagger}_{-\textbf{k},i}(x^0)=\beta^{r\dagger}_{-\textbf{k},i}\,e^{i\omega_{\bk,i}x^0}$. The operators
$\alpha^r_{\textbf{k},i}$ annihilate 
a field-mode of quantum numbers $\bk,r$ and energy $\omega_{\bk,i}=\sqrt{m_i^2+\bk^2}$.
They are defined by
\be
\label{masvac}
\alpha^r_{\textbf{k},i}|0\rangle_m\,=\,\beta^r_{\textbf{k},i}|0\rangle_m\,=\,0\,,\qquad i=1,2\,,
\ee
where $|0\rangle_m\equiv|0\rangle_1\otimes|0\rangle_2$ is the ``mass vacuum''. Similar relations hold for  $\beta^r_{\textbf{k},i}$. 

By imposing the canonical anti-commutation relations for the fields,
\be
\left\{\nu_i^\alpha(x),\nu_j^{\beta\dagger}(y)\right\}_{x^0=x^{0'}}\,=\,\delta^3{\left(\bx-\by\right)}\,\delta_{\alpha\beta}\,\delta_{ij}\,,\qquad \alpha,\beta=1,\dots,4\,,
\ee
we have
\be
\left\{\alpha^r_{\textbf{k},i},\alpha^{s\dagger}_{\textbf{q},j}
\right\}\,=\,\delta_{\bk\bq}\delta_{rs}\delta_{ij}\,,\qquad i,j=1,2\,,
\ee
and similarly for $\beta^r_{\bk,i}$, with all other anti-commutators
vanishing. 

For more details on the explicit form used for the spinors $u_{\textbf{k},i}^r$, $v_{-\textbf{k},i}^r$
and the Dirac $\gamma$-matrices, we remand to~\cite{BV95}. Here,  it is enough to write down the following
orthonormality and completeness relations 
\begin{eqnarray}
\label{ort}
&u_{\bk,i}^{r\dagger}u_{\bk,i}^s\,=\,v_{\bk,i}^{r\dagger}v_{\bk,i}^s\,=\,\delta_{rs}\,,\\[2mm]
&u_{\bk,i}^{r\dagger}v_{-\bk,i}^s\,=\,v_{-\bk,i}^{r\dagger}u_{\bk,i}^s=0\,,
\\[2mm]
\label{comp}
&\sum_{r}\left(u_{\bk,i}^r u_{\bk,i}^{r\dagger}\,+\,v_{-\bk,i}^{r}v_{-\bk,i}^{r\dagger}
\right)= \mathds{1}\,.
\end{eqnarray}

For reasons that will appear clear below, in Eq.~\eqref{fe}
we have considered the finite-volume expansion of free fields. 
Formally, the infinite-volume limit is obtained
by implementing the standard prescription 
\begin{eqnarray}
\sqrt{V} {\alpha}_{\bk} & \rightarrow & (2 \pi)^{{3}/{2}}\, {\alpha}_\bk\,,  \\[2mm]
\frac{1}{V}\sum_{\bk} & \rightarrow & \frac{1}{(2 \pi)^3}\, \int\hspace{0.1mm} d^3\textbf{k}\,, \\[2mm]
\frac{V \delta_{\bk \bq}}{(2 \pi)^3} & \rightarrow & \delta^3(\bk-\bq)\,.
\end{eqnarray}

Let us now turn to the expansion of flavor fields.
For this purpose, it is convenient to express the transformations~\eqref{Pontecf}
in terms of the mixing generator~\cite{BV95}
\be
\label{Gene}
G_\theta(x^0)\,=\,\exp\lf\{\theta\int d^3\bx\,\lf[{\nu}^\dagger_1(x){\nu}_2(x)-{\nu}^\dagger_2(x){\nu}_1(x)\ri]\ri\}\,,
\ee
which gives
\begin{equation}
\label{bogolgene}
{\nu}^\alpha_\ell(x)\,=\,{G}_\theta^{-1}(x^0)\,{\nu}^\alpha_i(x)\,{G_\theta}(x^0)\,,\qquad (\ell,i)=(e,1), (\mu,2)\,.
\end{equation}

At finite volume $V$, it is straightforward to see that this is a unitary operator, i.e. $G^{-1}_\theta(x^0)=G_{-\theta}(x^0)=G^\dagger_\theta(x^0)$, which preserves the canonical anitcommutators. Furthermore, it provides us with the map between the Hilbert space for free fields $\mathcal{H}_m$ and that for mixed fields $\mathcal{H}_f$. In particular, the vacuum for flavor fields (``flavor vacuum'') $|0(x^0)\rangle_f$ is given by
\be
\label{flv}
|0(x^0)\rangle_f=G_\theta^{-1}(x^0)|0\rangle_m\,.
\ee
Following the convention of~\cite{BV95}, we shall denote
by $|0\rangle_f$ the flavor vacuum at $x^0=0$. 

While being unitary for finite $V$, in the infinite-volume limit
(i.e., for systems with infinite degrees of freedom like fields)
$G_\theta(x^0)$ exhibits a non-unitary nature, which results
into the orthogonality of flavor and mass vacua~\cite{BV95,JiMi}
\be
_f\langle0(x^0)|0\rangle_m=0,\,\,\,\, \forall x^0\,.
\ee 
Likewise, flavor vacua at different times are orthogonal to each other~\cite{Terra}, i.e. $_f\langle0(x^{0'})|0(x^0)\rangle_f=0, \forall x^{0'}\neq x^{0}$. We stress that these are distinctive QFT features that are missing in QM, due to 
the validity of the Stone-von Neumann theorem~\cite{Stone,VN}. 
To avoid any technical issue, in what follows we shall perform all computations at finite volume and only at the end we consider the $V\rightarrow\infty$ limit. 

By letting the generator $G_\theta(x^0)$ act on $\nu_i(x)$ in Eq.~\eqref{bogolgene}, the flavor fields take the form 
\be
\label{flfiexp}
\nu_\ell(x)\,=\,\frac{1}{\sqrt{V}}\sum_{\textbf{k},r}\left[u_{\textbf{k},i}^r\,\alpha^r_{\textbf{k},\nu_\ell}(x^0)\,+\,v_{-\textbf{k},i}^r\,\beta^{r\dagger}_{-\textbf{k},\nu_\ell}(x^0)
\right]e^{i\textbf{k}\cdot\textbf{x}}\,,
\ee
with $(\ell,i)=(e,1),(\mu,2)$. The flavor annihilators are given by
\begin{eqnarray}
\label{boga}
\alpha^r_{\bk,\nu_e}(x^0)&=&\cos\theta\,\alpha^r_{\bk,1}(x^0)\,+\,\sin\theta\sum_{s}\left[u^{r\dagger}_{\bk,1}u^s_{\bk,2}\, \alpha^s_{\bk,2}(x^0)\,+\, u^{r\dagger}_{\bk,1} v^s_{-\bk,2}\,\beta^{s\dagger}_{-\bk,2}(x^0)\right],\\[2mm]
\label{bognewa}
\alpha^r_{\bk,\nu_\mu}(x^0)&=&\cos\theta\,\alpha^r_{\bk,2}(x^0)\,-\,\sin\theta\sum_{s}\left[u^{r\dagger}_{\bk,2}u^s_{\bk,1}\, \alpha^s_{\bk,1}(x^0)\,+\, u^{r\dagger}_{\bk,2} v^s_{-\bk,1}\,\beta^{s\dagger}_{-\bk,1}(x^0)\right],\\[2mm]
\label{bogb}
\beta^r_{-\bk,\nu_e}(x^0)&=&\cos\theta\,\beta^r_{-\bk,1}(x^0)\,+\,\sin\theta\sum_{s}\left[ v^{s\dagger}_{-\bk,2}v^r_{-\bk,1}\, \beta^s_{-\bk,2}(x^0)
\,+\, u^{s\dagger}_{\bk,2}v^r_{-\bk,1}\,\alpha^{s\dagger}_{\bk,2}(x^0)
\right],\\[2mm]
\label{bognewb}
\beta^r_{-\bk,\nu_\mu}(x^0)&=&\cos\theta\,\beta^r_{-\bk,2}(x^0)\,-\,\sin\theta\sum_{s}\left[ v^{s\dagger}_{-\bk,1}v^r_{-\bk,2}\, \beta^s_{-\bk,1}(x^0)
\,+\, u^{s\dagger}_{\bk,1}v^r_{-\bk,2}\,\alpha^{s\dagger}_{\bk,1}(x^0)
\right].
\end{eqnarray}
Notice that these transformations can be inverted
by utilizing the symmetry $\nu_\ell(x)\leftrightarrow\nu_i(x)$ when $\theta\rightarrow-\theta$. The time-dependence of the flavor operators~\eqref{bognewa}-\eqref{bognewb} indicates that flavor fields are interacting fields. It must be noted that this interacting model can be solved exactly, without needing perturbation expansions~\cite{Jizba}.
Furthermore, we emphasize that the choice of the reference time $x^0=0$ at which the flavor vacuum~\eqref{flv} and the related particle states are defined is not unique. In principle, any other choice would be possible and equivalent, provided that the flavor states which are acted upon by the corresponding flavor operators~\eqref{bognewa}-\eqref{bognewb} are consistently evaluated at the same time as the operators themselves and the commutators are all considered at equal times~\cite{BV95}.

As remarked in~\cite{Fuji,BlasDiMa}, the field 
expansion~\eqref{flfiexp} relies on a special choice of the bases of spinors, namely those referring to the free field masses $m_1$ and $m_2$, respectively. However, it is always possible to perform a Bogoliubov transformation in order to expand the field operators in a different basis of spinors, referring to an arbitrarily chosen couple of mass parameters (for instance, the natural bases corresponding to the couple of masses $m_e$, $m_\mu$). The relevant point is that this transformation leaves measurable quantities (such as flavor charges and oscillation probabilities) invariant, as it should be. Therefore, in this sense flavor mixing can be reformulated as a gauge theory~\cite{BlasDiMa}.

Equations~\eqref{boga}-\eqref{bognewb} 
clearly show the non-trivial structure of flavor mixing in QFT: in fact, besides the standard Pontecorvo rotation, flavor annihilators also contain a Bogoliubov transformation (the terms in the brackets) arising from the products of (anti-)spinors with different masses. The presence of such extra terms lies at the heart of the QFT treatment of mixing, as they imply that flavor vacuum annihilators do not annihilate mass vacuum, and vice-versa. 

Without loss of generality, we can now select the reference frame such that
$\bk=\left(0,0,|\bk|\right)$. In this setting, only the products of wave functions with $r = s$ are non-vanishing, which allows us to rewrite Eqs.~\eqref{boga}-\eqref{bognewb} in the simpler form
\begin{eqnarray}
\label{boga2}
\alpha^r_{\bk,\nu_e}(x^0)&=&\cos\theta\,\alpha^r_{\bk,1}(x^0)\,+\,\sin\theta\left(|U_\bk|\,\alpha^r_{\bk,2}(x^0)\,+\,\epsilon^r |V_\bk|\,\beta^{r\dagger}_{-\bk,2}(x^0)
\right),\\[2mm]
\label{bogc2}
\alpha^r_{\bk,\nu_\mu}(x^0)&=&\cos\theta\,\alpha^r_{\bk,2}(x^0)\,-\,\sin\theta\left(|U_\bk|\,\alpha^r_{\bk,1}(x^0)\,-\,\epsilon^r |V_\bk|\,\beta^{r\dagger}_{-\bk,1}(x^0)
\right),\\[2mm]
\label{bogb2}
\beta^r_{-\bk,\nu_e}(x^0)&=&\cos\theta\,\beta^r_{-\bk,1}(x^0)\,+\,\sin\theta\left(|U_\bk|\,\beta^r_{-\bk,2}(x^0)\,-\,\epsilon^r |V_\bk|\,\alpha^{r\dagger}_{\bk,2}(x^0)
\right),\\[2mm]
\label{bogc3}
\beta^r_{-\bk,\nu_\mu}(x^0)&=&\cos\theta\,\beta^r_{-\bk,2}(x^0)\,-\,\sin\theta\left(|U_\bk|\,\beta^r_{-\bk,1}(x^0)\,+\,\epsilon^r |V_\bk|\,\alpha^{r\dagger}_{\bk,1}(x^0)
\right),
\end{eqnarray} 
where $\epsilon^r=(-1)^r$.
The Bogoliubov coefficients $|U_\bk|$ and $|V_\bk|$ are given by
\begin{eqnarray}
\label{U}
\nonumber
|U_\bk|&=&u_{\bk,i}^{r\dagger}\,u_{\bk,j}^r\,=\,v^{r\dagger}_{-\bk,i}\,v^r_{-\bk,j}\\[2mm]
&=&\frac{|\bk|^2+\left(\omega_{\bk,1}+m_1\right)\left(\omega_{\bk,2}+m_2\right)}{2\sqrt{\omega_{\bk,1}\omega_{\bk,2}\left(\omega_{\bk,1}+m_1\right)\left(\omega_{\bk,2}+m_2\right)}}\,,\\[4mm]
\nonumber
|V_\bk|&=&\epsilon^r\,u_{\bk,1}^{r\dagger}\,v^r_{-\bk,2}\,=\,-\epsilon^r\, u^{r\dagger}_{\bk,2}v_{-\bk,1}^r\\[2mm]
&=&\frac{\left(\omega_{\bk,1}+m_1\right)-\left(\omega_{\bk,2}+m_2\right)}{2\sqrt{\omega_{\bk,1}\omega_{\bk,2}\left(\omega_{\bk,1}+m_1\right)\left(\omega_{\bk,2}+m_2\right)}}\,|\bk|\,,
\label{V}
\end{eqnarray}
with $i,j=1,2$, $i\neq j$ (similar relations hold for $\bk\rightarrow-\bk$). 
It is a matter of calculations to show that
\be
|U_{\bk}|^2\,+\,|V_{\bk}|^2\,=\,1\,,
\ee
which ensures that flavor ladder operators still satisfy canonical anti-commutation relations (at equal times). 

For our next purposes, we note that the following
relations hold true in the reference frame where $\bk=\left(0,0,|\bk|\right)$: 
\begin{eqnarray}
\label{prop1}
\bar u_{\bk,1}^r |U_\bk|\,-\,\epsilon^r\, \bar v^r_{-\bk,1} |V_{\bk}|&=&\bar u_{\bk,2}^r\,,\\[2mm]
\bar u_{\bk,1}^r |V_\bk|\,+\,\epsilon^r\, \bar v^r_{-\bk,1} |U_{\bk}|&=&\epsilon^r\,\bar v_{-\bk,2}^r\,,\\[2mm]
\label{prop2}
\label{prop3}
v^r_{\bk,1}|U_{\bk}|\,-\,\epsilon^r\, u^r_{-\bk,1} |V_{\bk}|&=&v_{\bk,2}^r\,,\\[2mm]
\label{prop4}
v^r_{\bk,1}|V_{\bk}|\,+\,\epsilon^r\,u^r_{-\bk,1} |U_{\bk}|&=&\epsilon^r\,u_{-\bk,2}^r\,,
\end{eqnarray}
where $\bar u_{\bk,i}^r=u_{\bk,i}^{r\dagger}\,\gamma_0$ and $\bar v_{-\bk,i}^r=v_{-\bk,i}^{r\dagger}\,\gamma_0$, $i=1,2$. Furthermore, we have
\begin{eqnarray}
\label{prop5}
\bar u_{\bk,2}^r\,|U_\bk|\,+\,\epsilon^r\,\bar v^r_{-\bk,2}\,|V_\bk|&=&\bar u_{\bk,1}^r\,,\\[2mm]
\label{prop6}
v_{\bk,2}^r\,|U_\bk|\,+\,\epsilon^r\,u^r_{-\bk,2}\,|V_\bk|&=&v_{\bk,1}^r\,.
\end{eqnarray}

QFT neutrino states are now defined by the action
of flavor creation operators on the flavor vacuum $|0\rangle_f$ at $x^0=0$, i.e.~\cite{BV95}
\be
\label{fqfts}
|\nu^r_{\bk,\ell}\rangle\,\equiv\,\alpha^{r\dagger}_{\bk,\nu_\ell}(0)|0\rangle_f\,,\qquad \ell=e,\mu\,.
\ee
As usual, the time evolution of these states is ruled by
$|\nu^r_{\bk,\ell}(x^0)\rangle=e^{iH_0x^0}\alpha^{r\dagger}_{\bk,\nu_\ell}(0)|0\rangle_f$. 

\smallskip

It should be noticed that the QFT formalism
described above reproduces the standard QM Pontecorvo treatment in the ultra-relativistic limit $\frac{\Delta m}{|\bk|}=\frac{|m_2-m_1|}{{|\bk|}}\ll1$. To this end, we observe that the Bogoliubov coefficients $|U_\bk|$ and $|V_{\bk}|$ at the first order in $\mathcal{O}\left(\frac{\Delta m}{2|\bk|}\right)$ take the form 
\begin{subequations}
\label{relat}
\begin{align}
\label{relata}
|U_\bk|&\,\approx\, 1\,,\\[2mm]
\label{relatb}
|V_\bk|&\,\approx\, \frac{\Delta m}{2|\bk|}\,,
\end{align}
\end{subequations}
from which we get $|U_{\bk}|\rightarrow1, |V_{\bk}|\rightarrow0$ as $\frac{\Delta m}{|\bk|}\rightarrow0$. Inserting these values into the Bogoliubov transformations~\eqref{boga2}-\eqref{bogc3}, it is straightforward to see that Pontecorvo rotations for flavor annihilators are trivially recovered, which 
entails that flavor and mass vacua 
are equivalent to each other in this limit. In turn, this implies
that the definition~\eqref{fqfts} of exact QFT flavor states
reduces to the usual Pontecorvo definition~\eqref{Pontec}.

The above considerations provide us with the basic ingredients to analyze the effects of neutrino mixing in a self-consistent field theoretical way. In particular, we shall resort to the field expansions~\eqref{flfiexp} and the Bogoliubov transformations~\eqref{boga2}-\eqref{bogc3} to compute the transition amplitude
for the neutron $\beta$-decay.

\section{Neutron $\beta$-decay with Pontecorvo and mass states for neutrinos}
\label{Wip}
In this Section we consider the weak decay
\be
\label{ndec}
a)\,\,\,n\,\rightarrow\, p \,+\,e^-\,+\,\bar\nu_e\,,
\ee
as paradigmatic process to check the consistency of neutrino mixing with the conservation of 
lepton charge in the (tree-level) interaction vertex\footnote{Although our considerations are specific for the $\beta$-decay process, results and conclusions are more general and can be extended to all weak interactions involving neutrinos.}. We stress that
this symmetry is familiar to any physics student
from the most common muon decay mode $\mu^-\,\rightarrow\,e^{-}\bar\nu_e\nu_\mu$, which would be a direct conversion $\mu^-\,\rightarrow\, e^-$ without the introduction of neutrinos to both balance
lepton flavor and explain the $3$-body spectrum of electron energy. 

In parallel to the interaction~\eqref{ndec}, we also consider 
the flavor-violating process
\be
\label{ndecmu}
b)\,\,\,n\,\rightarrow\, p \,+\,e^-\,+\,\bar\nu_\mu\,,
\ee
which is in principle allowed because of neutrino oscillations. Nevertheless, for time intervals sufficiently longer than the neutron
lifetime $\tau_n$, but much shorter compared to the characteristic  oscillation time $T_{osc}$, the tree-level amplitude for this process is expected to vanish for consistency with lepton charge conservation. We shall refer to this time scale as ``short-time'' limit. Of course, given the experimental values of $\tau_n$ and $T_{osc}$, 
such interval is well defined. 

In the scattering theory for finite-range potentials, 
it is typically assumed that the interaction Hamiltonian $H_{int}(x)$
can be switched off far before and after the interaction, i.e. $\lim_{x^0_{in}\rightarrow-\infty}H_{int}(x)=\lim_{x^0_{out}\rightarrow+\infty}H_{int}(x)=0$ (adiabatic approximation)~\cite{Itz}, so that 
initial and final states can be 
represented as eigenstates of the free hamiltonian $H_0$.
However, in the presence of mixed neutrinos, such an approximation
fails, since neutrino fields cannot be defined as asymptotic operators acting on the mass vacuum. As discussed in Sec.~\ref{QFTFo}, this
follows from the fact that Fock spaces for flavor and mass fields
are unitarily inequivalent to each other. The supposed ambiguity has been successfully addressed in~\cite{CYL}, showing that the long-time limit of the amplitudes~\eqref{ndec} and~\eqref{ndecmu} is actually 
well-defined when working with exact QFT states. Furthermore, this limit turns out to be consistent with SM predictions in the relativistic approximation, which corroborates the physical rigor of the definition of QFT neutrino states. 

For our purposes of computing the short-time limit of 
the decay amplitudes~\eqref{ndec} and~\eqref{ndecmu}, 
we consider the perturbation theory at the first order.  We then set the
integration limits $x^0_{in}$, $x^0_{out}$ such that
$\Delta x^0=x^0_{out}-x^0_{in}\ll T_{osc}$, as discussed above. In the interaction picture, 
denoting the ingoing and outgoing states by $|\psi_i\rangle$ and $|\psi_f\rangle$, respectively, the probability amplitude is given by
$\langle e^{iH_0x^0}\psi_f|U_I(x^0)|\psi_i\rangle$,
where the time evolution operator $U_I(x^0)$ to the leading order is 
\be
U_I(x^0)\simeq 1-i\int_{0}^{x^0}dx^{0'}\,H_{int}(x^{0'})\,.
\ee 
Here, $H_{int}(x^{0})=e^{iH_0x^0}\,H_{int}\,e^{-iH_0x^0}$ is the
interaction Hamiltonian in the interaction picture.

The effective Lagrangian for the interaction~\eqref{ndec} is given by~\cite{Cli}
\be
\label{Hint}
\mathscr{L}_{int}(x)\,=\, -\frac{G_F}{\sqrt{2}}\left[\bar e(x)\hspace{0.2mm}\gamma^\mu(\mathds{1}-\gamma^5)\nu_e(x)
\right]\left[V_{ud}\, \bar q_u(x) \gamma_\mu\hspace{0.2mm} (f\,-\,\gamma^5 g)\hspace{0.2mm} q_d(x)
\right]\,,
\ee
where $G_F$ is the Fermi constant, $V_{ud}$ is the CKM matrix element
describing the coupling between the up and down quarks, while $f,g$ are the form factors. The electron, neutrino, up and down quark fields
have been denoted by $e(x)$, $\nu_e(x)$, $q_u(x)$ and $q_d(x)$, respectively.  As well known, the projection operator $P_L=1/2\left(\mathds{1}-\gamma_5\right)$ appears because
charged-current in weak interactions only couple to left-chiral fermions. 
Furthermore, we have omitted the hermitian conjugate in Eq.~\eqref{Hint}, since it gives a vanishing contribution to the transition amplitude. 

The above expression of $\mathscr{L}_{int}$ is 
justified within the framework of (current-current interaction) Fermi theory. Clearly, the actual Lagrangian is more complicated than Eq.~\eqref{Hint} (see~\cite{Act}). However, 
such additional complications will be mostly irrelevant
for our following discussion and can be safely overlooked. 

Based on the above considerations, let us compute the amplitude 
for the decay channels $a)$ and $b)$. We perform calculations
by resorting first to Pontecorvo states $|\nu_{\bk,\ell}\rangle_P$
in Eqs.~\eqref{Pontec} and then to mass states $|\nu_{\bk,i}\rangle$ as  neutrino fundamental representation. 
At a later stage, we shall focus on the short-time limit of the resulting amplitudes, discussing the regime of their validity.

\subsection{Pontecorvo states}
As a first step, we consider the process $a)$. 
The transition amplitude at tree level takes the form
\be
A^{a)}_{P}\,=\,_P\langle\bar\nu^r_{\bk,e}, e^s_{\bq},\hspace{0.4mm} p^{\sigma_p}_{\bp_p} | 
\left[-i\int_{x^{0}_{in}}^{x^{0}_{out}}d^4x\, \mathscr{L}_{int}(x)
\right] |n^{\sigma_n}_{\bp_n}\rangle\,,
\label{Aa}
\ee
where we have denoted by $\bp_n $ ($\sigma_n$),  
$\bp_p$ ($\sigma_p$) and $\bq$ ($s$) the momentum (helicity)
of the neutron, proton and electron, respectively, 
and utilized Pontecorvo state for the outgoing (anti-)neutrino.  
To simplify the notation, we have omitted
the time-dependence of particle states. Clearly, in the standard
scattering theory in the interaction picture, the ingoing/outgoing states
must be evaluated at $x^0=x^0_{in}$ and $x^0=x^0_{out}$, respectively\footnote{As observed in~\cite{CapW}, the amplitude for the case of the exact QFT neutrino states should be consistently defined as $\langle \psi_\ell(x^0_{out})|e^{-iH\left(x_{out}^0-x_{in}^0\right)}|\psi_\ell(x^0_{in})\rangle=\langle \psi_\ell(x^0_{in})|U_I(x^0_{out},x_{in}^0)|\psi_\ell(x^0_{in})\rangle$, due to the orthogonality of the Hilbert spaces at different times (see Sec.~\ref{QFTFo}). This will be explicitly taken into account in Sec.~\ref{ExQFTs}.}. 

Now, by plugging Eq.~\eqref{Hint} into~\eqref{Aa}, the term involving the expectation value
of the quark up and down fields
can be easily computed to give
\begin{eqnarray}
\non
\label{prot}
\langle p^{\sigma_p}_{\bp_p} | V_{ud}\, \bar q_u(x) \gamma_\mu\hspace{0.2mm} (f\,-\,\gamma^5 g)\hspace{0.2mm} q_d(x) |n^{\sigma_n}_{\bp_n}\rangle &\simeq& V_{ud}\, \bar u_{\bp_p}^{\sigma_p}\hspace{0.3mm}\gamma_\mu (f\,-\,\gamma^5 g)\hspace{0.3mm} u_{\bp_n}^{\sigma_n}\\[2mm]
&&\times\,e^{-i(\omega_{\bp_p}x_{out}^0-\omega_{\bp_n}x^0_{in})}\,e^{-i\left[\left(\omega_{\bp_n}-\omega_{\bp_p}\right)\hspace{0.2mm}x^0-\left(\bp_n-\bp_p\right)\cdot\bx\right]}\,,
\end{eqnarray}
up to a normalization factor.
Here we have denoted by $\omega_{\bp_n}$ and $\omega_{\bp_p}$ the 
energy of the neutron and proton, respectively. 

Concerning the lepton current, it is convenient to compute separately the terms involving the electron and neutrino fields. For the former we have
\be
\label{elf}
\langle e^s_\bq | \bar e(x) | 0\rangle_e\, \simeq\, \bar u_{\bq,e}^s\,e^{-i\bq\cdot\bx}\,e^{-i\omega_\bq^e(x^0_{out}-x^0)}\,,
\ee
where $\omega_\bq^e$ is the electron energy. 
On the other hand,  we get for the neutrino field
\be
\label{ampP}
_P\langle\bar\nu^r_{\bk,e} | \nu_e(x) | 0\rangle_m\, \,=\, \cos^2\theta\, _P\langle\bar\nu^r_{\bk,1} | \nu_1(x) | 0\rangle_m\,+\,\sin^2\theta\, _P\langle\bar\nu^r_{\bk,2} | \nu_2(x) | 0\rangle_m\,,
\ee
where $|0\rangle_m$ is the vacuum for the massive neutrino fields
$\nu_i(x)$, $i=1,2$ (see Eq.~\eqref{masvac}). 
Notice that we have used the mixing transformations~\eqref{Ponteca} and~\eqref{Pontecfa} for 
the neutrino state and field, respectively. 

Equation~\eqref{ampP} can be
further manipulated to give
\be
\label{expneu}
_P\langle\bar\nu^r_{\bk,e} | \nu_e(x) | 0\rangle_m\,\simeq\,e^{-i\bk\cdot\bx}\left[\cos^2\theta\, v^r_{\bk,1}\, e^{-i\omega_{\bk,1}\left(x^0_{out}-x^0\right)}
\,+\,\sin^2\theta\, v^r_{\bk,2}\, e^{-i\omega_{\bk,2}\left(x^0_{out}-x^0\right)}
\right],
\ee
where we have still omitted the normalization. 

Combining Eqs.~\eqref{prot},~\eqref{elf} and~\eqref{expneu}, the amplitude~\eqref{Aa} becomes
\begin{eqnarray}
\nonumber
A^{a)}_{P}&=&i\,\mathcal{N}\hspace{0.1mm}G_F\, 
\delta^{3}\left(\bp_n-\bp_p-\bq-\bk\right)e^{-i\omega_\bq^ex^0_{out}}\,\bar{u}^s_{\bq,e}\,\gamma^{\mu}(\mathds{1}-\gamma^5)\\[2mm]
\nonumber
&&\times\int_{x_{in}^0}^{x_{out}^0}\,dx^0\left[\cos^2\theta\,v^r_{\bk,1}\,e^{-i\omega_{\bk,1}x^0_{out}}\,e^{-i\left(\omega_{\bp_n}-\omega_{\bp_p}-\omega_{\bq}^e-\omega_{\bk,1}\right)x^0}\right.\\[2mm]\nonumber
&&\left.+\,\sin^2\theta\,v^r_{\bk,2}\,e^{-i\omega_{\bk,2}x^0_{out}}\,e^{-i\left(\omega_{\bp_n}-\omega_{\bp_p}-\omega_{\bq}^e-\omega_{\bk,2}\right)x^0}
\right]\\[2mm] 
&&\times\,V_{ud}\, \bar u_{\bp_p}^{\sigma_p}\hspace{0.3mm}\gamma_\mu (f\,-\,\gamma^5 g)\hspace{0.3mm} u_{\bp_n}^{\sigma_n}\,e^{-i(\omega_{\bp_p}x_{out}^0-\omega_{\bp_n}x^0_{in})}\,,
\label{aa}
\end{eqnarray}
where we have encompassed 
all the factors omitted above in the normalization $\mathcal{N}=\left[{\sqrt{2}\left(2\pi\right)^3}\right]^{-1}$. 

In a similar fashion, if we consider the decay channel $b)$ in Eq.~\eqref{ndecmu}, the amplitude reads
\be
A^{b)}_{P}\,=\,_P\langle\bar\nu^r_{\bk,\mu}, e^s_{\bq},\hspace{0.4mm} p^{\sigma_p}_{\bp_p} | 
\left[-i\int_{x^{0}_{in}}^{x^{0}_{out}}d^4x\, \mathscr{L}_{int}(x)
\right] |n^{\sigma_n}_{\bp_n}\rangle\,.
\label{Amu}
\ee
In this case, the expectation value involving the neutrino field
is
\be
\label{numu}
_P\langle\bar\nu^r_{\bk,\mu} | \nu_e(x) | 0\rangle_m\,\simeq\, 
e^{-i\bk\cdot\bx}\,\sin\theta\cos\theta\left[v^r_{\bk,2}\,e^{-i\omega_{\bk,2}\left(x^0_{out}-x^0\right)}\,-\,
v^r_{\bk,1}\,e^{-i\omega_{\bk,1}\left(x^0_{out}-x^0\right)}
\right],
\ee
where we have resorted to the transformation~\eqref{Pontecb}
for the state $|\bar\nu_{\bk,\mu}\rangle_P$. 

From Eq.~\eqref{numu}, we get
\begin{eqnarray}
\nonumber
A^{b)}_{P}&=&i\,\mathcal{N}\hspace{0.1mm}G_F\,\sin\theta\cos\theta\,
\delta^{3}\left(\bp_n-\bp_p-\bq-\bk\right)e^{-i\omega_\bq^ex^0_{out}}\,\bar{u}^s_{\bq,e}\,\gamma^{\mu}(\mathds{1}-\gamma^5)\\[2mm]
\nonumber
&&\times\int_{x_{in}^0}^{x_{out}^0}\,dx^0\, \left[v^r_{\bk,2}\,e^{-i\omega_{\bk,2}x^0_{out}}\,e^{-i\left(\omega_{\bp_n}-\omega_{\bp_p}-\omega_{\bq}^e-\omega_{\bk,2}\right)x^0}\right.\\[2mm]
\nonumber
&&\left.-\,v^r_{\bk,1}\,e^{-i\omega_{\bk,1}x^0_{out}}\,e^{-i\left(\omega_{\bp_n}-\omega_{\bp_p}-\omega_{\bq}^e-\omega_{\bk,1}\right)x^0}
\right]\\[2mm]
&&\times\,V_{ud}\, \bar u_{\bp_p}^{\sigma_p}\hspace{0.3mm}\gamma_\mu (f\,-\,\gamma^5 g)\hspace{0.3mm} u_{\bp_n}^{\sigma_n}\,e^{-i(\omega_{\bp_p}x_{out}^0-\omega_{\bp_n}x^0_{in})}\,.
\label{ab}
\end{eqnarray}
For vanishing mixing (i.e. $\theta\rightarrow0$), 
this expression is identically zero, as expected 
in the absence of flavor oscillations. 

\subsection{Short-Time Limit}
\label{Aistl}
Let us now use Eqs.~\eqref{aa} and \eqref{ab} to analyze lepton charge  conservation in the tree-level interaction vertex
for the processes $n\,\rightarrow\, p \,+\,e^-\,+\,\bar\nu_e$ and $n\,\rightarrow\, p \,+\,e^-\,+\,\bar\nu_\mu$. For this purpose, 
we study the structure of $A_P^{a)}$ and $A_P^{b)}$ in the
short-time approximation. We 
set the limits of integration
as $x^0_{in}=-\Delta t/2$ and $x^0_{out}=\Delta t/2$, such that
$\Delta t$ is sufficiently longer than the neutron lifetime $\tau_n$ to ensure (on average) the neutron decay, but 
shorter than the characteristic neutrino oscillation time $T_{osc}$, 
in compliance with the discussion below Eq.~\eqref{ndecmu}.
Notice that a similar assumption has been considered in~\cite{Nishi}
in the context of the pion decay.

With reference to the process $a)$, the transition amplitude~\eqref{aa} becomes
\begin{eqnarray}
\nonumber
A^{a)}_{P}&=&i\,\mathcal{N}\hspace{0.1mm}G_F\, 
\delta^{3}\left(\bp_n-\bp_p-\bq-\bk\right)e^{-i\omega_\bq^e\Delta t/2}\,\bar{u}^s_{\bq,e}\,\gamma^{\mu}(\mathds{1}-\gamma^5)\\[2mm]
\nonumber
&&\times\int_{-\Delta t/2}^{\Delta t/2}\,dx^0\left[\cos^2\theta\,v^r_{\bk,1}\,e^{-i\omega_{\bk,1}\Delta t/2}\,e^{-i\left(\omega_{\bp_n}-\omega_{\bp_p}-\omega_{\bq}^e-\omega_{\bk,1}\right)x^0}\right.\\[2mm]
&&\nonumber
\left.+\,\sin^2\theta\,v^r_{\bk,2}\,e^{-i\omega_{\bk,2}\Delta t/2}\,e^{-i\left(\omega_{\bp_n}-\omega_{\bp_p}-\omega_{\bq}^e-\omega_{\bk,2}\right)x^0}
\right] \\[2mm]
&&\times\,V_{ud}\, \bar u_{\bp_p}^{\sigma_p}\hspace{0.3mm}\gamma_\mu (f\,-\,\gamma^5 g)\hspace{0.3mm} u_{\bp_n}^{\sigma_n}\,e^{-i(\omega_{\bp_p}+\omega_{\bp_n})\Delta t/2}\,.
\label{aasl}
\end{eqnarray}
By integrating over $x^0$, we then get
\begin{eqnarray}
\nonumber
A^{a)}_{P}&=&2i\,\mathcal{N}\hspace{0.1mm}G_F\, 
\delta^{3}\left(\bp_n-\bp_p-\bq-\bk\right)e^{-i\omega_\bq^e\Delta t/2}\,\bar{u}^s_{\bq,e}\,\gamma^{\mu}(\mathds{1}-\gamma^5)\\[2mm]
\nonumber
&&\times\left\{\cos^2\theta\,v^r_{\bk,1}\,e^{-i\omega_{\bk,1}\Delta t/2}\
\frac{\sin\left[\left(\omega_{\bp_n}-\omega_{\bp_p}-\omega_{\bq}^e-\omega_{\bk,1}
\right)\Delta t/2
\right]}{\omega_{\bp_n}-\omega_{\bp_p}-\omega_{\bq}^e-\omega_{\bk,1}}\right.\\[2mm]\nonumber
&&\left.\,+\sin^2\theta\,v^r_{\bk,2}\,e^{-i\omega_{\bk,2}\Delta t/2}\,
\frac{\sin\left[\left(\omega_{\bp_n}-\omega_{\bp_p}-\omega_{\bq}^e-\omega_{\bk,2}
\right)\Delta t/2
\right]}{\omega_{\bp_n}-\omega_{\bp_p}-\omega_{\bq}^e-\omega_{\bk,2}}
\right\}\\[3mm]
&&\times\,V_{ud}\, \bar u_{\bp_p}^{\sigma_p}\hspace{0.3mm}\gamma_\mu (f\,-\,\gamma^5 g)\hspace{0.3mm} u_{\bp_n}^{\sigma_n}\,e^{-i(\omega_{\bp_p}+\omega_{\bp_n})\Delta t/2}\,.
\label{43}
\end{eqnarray}

A comment is now in order: in the limit of large $\Delta t$, 
the $\sin(\tilde\omega\Delta t)/\tilde\omega$ factors 
in the above relation become the usual energy-conserving
Dirac $\delta$ functions. Here we have used the shorthand notation
$\tilde\omega_i=\omega_{\bp_n}-\omega_{\bp_p}-\omega_{\bq}^e-\omega_{\bk,i}$, $i=1,2$.
On the other hand, the apparent energy fluctuations for finite time intervals can be understood in terms of (and are constrained by) the time-energy uncertainty relation for neutrino oscillations, with
$\Delta t$ being the interval defined above. 
This scenario has been rigorously analyzed in both perturbative QM~\cite{BilPhys} and QFT~\cite{SmaldFl}, where it has been shown that the 
Mandelstam-Tamm version of time-energy uncertainty relations
for neutrino oscillations can be cast in the form of flavor-energy uncertainty relations. In other terms, neutrino flavor charges obtained via Noether's theorem and energy turn out to be incompatible
observables in the usual quantum mechanical sense. 
Flavor-energy uncertainty relations 
have been recently studied also in stationary curved spacetime~\cite{BlasSma}.

\smallskip

Let us now consider Eq.~\eqref{43} in the short-time limit. Clearly, the dominant contributions in this regime are those
for which $\tilde\omega_i\approx0$. At the leading order, we obtain
\begin{eqnarray}
\nonumber
A^{a)}_{P}	&\simeq&i\,\mathcal{N}\hspace{0.1mm}G_F\, 
\delta^{3}\left(\bp_n-\bp_p-\bq-\bk\right)\Delta t\,\bar{u}^s_{\bq,e}\,\gamma^{\mu}(\mathds{1}-\gamma^5)\,
\left(\cos^2\theta\,v^r_{\bk,1}\,+
\sin^2\theta\,v^r_{\bk,2}
\right)\\[2mm]
&&\times\,V_{ud}\, \bar u_{\bp_p}^{\sigma_p}\hspace{0.3mm}\gamma_\mu (f\,-\,\gamma^5 g)\hspace{0.3mm} u_{\bp_n}^{\sigma_n}\,.
\label{App}
\end{eqnarray}

Since observed neutrinos are relativistic, it is convenient
to study the result~\eqref{App} in the relativistic limit. This approximation 
will also be useful for a more direct comparison 
with the analysis of the next Section. 
In this regard, we use the identity~\eqref{prop3} 
and rewrite $A^{a)}_{P}$ in the form
\begin{eqnarray}
\nonumber
A^{a)}_{P}	&\simeq&i\,\mathcal{N}\hspace{0.1mm}G_F\, 
\delta^{3}\left(\bp_n-\bp_p-\bq-\bk\right)\Delta t
\,\bar{u}^s_{\bq,e}\,\gamma^{\mu}(\mathds{1}-\gamma^5)\,
\left\{v^r_{\bk,1}\left[1-\sin^2\theta\left(1-|U_\bk|\right)\right]
-\sin^2\theta\,\epsilon^r\,u^r_{-\bk,1}\,|V_\bk|
\right\}\\[2mm]
&&\times\,V_{ud}\, \bar u_{\bp_p}^{\sigma_p}\hspace{0.3mm}\gamma_\mu (f\,-\,\gamma^5 g)\hspace{0.3mm} u_{\bp_n}^{\sigma_n}\,.
\label{Appbis}
\end{eqnarray}
By means of Eqs.~\eqref{relat}, we obtain to the first order in $\mathcal{O}\left(\frac{\Delta m}{2|\bk|}\right)$
\begin{eqnarray}
\nonumber
A^{a)}_{P}	&\simeq&i\,\mathcal{N}\hspace{0.1mm}G_F\, 
\delta^{3}\left(\bp_n-\bp_p-\bq-\bk\right)\Delta t
\,\bar{u}^s_{\bq,e}\,\gamma^{\mu}(\mathds{1}-\gamma^5)\,
\left(v^r_{\bk,1}\, -\,\sin^2\theta\,\frac{\Delta m}{2|\bk|}
\epsilon^r\,u^r_{-\bk,1}\right)\\[2mm]
&&\times\,V_{ud}\, \bar u_{\bp_p}^{\sigma_p}\hspace{0.3mm}\gamma_\mu (f\,-\,\gamma^5 g)\hspace{0.3mm} u_{\bp_n}^{\sigma_n}\,, 
\end{eqnarray}
which becomes in the ultarelativistic limit (i.e. for $\frac{\Delta m}{|\bk|}\rightarrow0)$
\begin{eqnarray}
\nonumber
A^{a)}_{P}	&\simeq&i\,\mathcal{N}\hspace{0.1mm}G_F\, 
\delta^{3}\left(\bp_n-\bp_p-\bq-\bk\right)\Delta t
\,\bar{u}^s_{\bq,e}\,\gamma^{\mu}(\mathds{1}-\gamma^5)\,
v^r_{\bk,1}\\[2mm] 
&&\times\,
V_{ud}\, \bar u_{\bp_p}^{\sigma_p}\hspace{0.3mm}\gamma_\mu (f\,-\,\gamma^5 g)\hspace{0.3mm} u_{\bp_n}^{\sigma_n}\,. 
\label{agSM}
\end{eqnarray}
Notice that this expression exhibits the same structure as the amplitude
for the emission of a free (anti-)neutrino with mass $m_1$. 
Further discussion on the meaning of this result will be 
given below.

We now consider the short-time approximation of the amplitude~\eqref{ab}.
Following the same steps as above, we are led to
\begin{eqnarray}
\label{absl}
\nonumber
A^{b)}_{P}&\simeq&i\,\mathcal{N}\hspace{0.1mm}G_F\,\sin\theta\cos\theta\,
\delta^{3}\left(\bp_n-\bp_p-\bq-\bk\right)\Delta t\,\bar{u}^s_{\bq,e}\,\gamma^{\mu}(\mathds{1}-\gamma^5)\left(v_{\bk,2}^r\,-\,v^r_{\bk,1}\right)\\[2mm]
&&\times\,V_{ud}\, \bar u_{\bp_p}^{\sigma_p}\hspace{0.3mm}\gamma_\mu (f\,-\,\gamma^5 g)\hspace{0.3mm} u_{\bp_n}^{\sigma_n}\,,
\end{eqnarray}
which is evidently non-vanishing, since $v_{\bk,2}^r\neq v^r_{\bk,1}$ for $m_2\neq m_1$. 

Again, it is interesting to consider the relativistic limit. 
By using Eqs.~\eqref{prop3} and~\eqref{relat}, we obtain after some algebra
\begin{eqnarray}
\label{re2}
\nonumber
A^{b)}_{P}&\simeq&-i\,\mathcal{N}\hspace{0.1mm}G_F\,\sin\theta\cos\theta\,
\delta^{3}\left(\bp_n-\bp_p-\bq-\bk\right)\Delta t\,\frac{\Delta m}{2|\bk|}\,\bar{u}^s_{\bq,e}\,\gamma^{\mu}(\mathds{1}-\gamma^5)\,\epsilon^r\,u^r_{-\bk,1}\,,\\[2mm]
&&\times\,V_{ud}\, \bar u_{\bp_p}^{\sigma_p}\hspace{0.3mm}\gamma_\mu (f\,-\,\gamma^5 g)\hspace{0.3mm} u_{\bp_n}^{\sigma_n}\,.
\end{eqnarray}
It must be emphasized that Eq.~\eqref{absl} (or, equivalently, Eq.~\eqref{re2}) signals a clear violation of lepton charge
in the tree-level interaction vertex, as it states that 
the flavor-violating process $n\,\rightarrow\, p \,+\,e^-\,+\,\bar\nu_\mu$ 
has a nonzero probability even for intervals much shorter 
than the characteristic neutrino oscillation time.
In the next Section, we will show that the origin of 
this flaw can be traced back to the incorrect definition
of Pontecorvo states as eigenstates of the flavor charge in QFT. As further evidence of this, from Eq.~\eqref{re2} we see that the correct vanishing amplitude is recovered
in the ultra-relativistic limit $\frac{\Delta m}{|\bk|}\rightarrow0$, 
where, in fact, Pontecorvo states well-approximate exact QFT states (see the discussion below Eqs.~\eqref{fqfts}).

\subsection{Mass states}
Proceeding in the same way as done in the previous subsection,
we now compute the neutron $\beta$-decay amplitude
by using the mass eigenstates $|\nu^r_{\bk,i}\rangle$, $i=1,2$, 
as fundamental representation for neutrinos. This amounts to consider the two processes 
\be
\label{masproc}
n\,\rightarrow\, p \,+\,e^-\,+\,\bar\nu_i\,,\qquad i=1,2\,.
\ee 
In this context, we observe
that Eq.~\eqref{Aa} (or, equivalently, Eq.~\eqref{Amu}) 
must be rewritten as
\be
A_i=\langle\bar\nu^r_{\bk,i}, e^s_{\bq},\hspace{0.4mm} p^{\sigma_p}_{\bp_p} | 
\left[-i\int_{x^{0}_{in}}^{x^{0}_{out}}d^4x\, \mathscr{L}_{int}(x)
\right] |n^{\sigma_n}_{\bp_n}\rangle\,,\qquad i=1,2\,,
\label{iamp}
\ee
where the subscript $i$ refers to the $i^{th}$ mass
state.

The amplitude~\eqref{iamp} can be straightforwardly 
derived from Eq.~\eqref{aa} by using 
the transformation~\eqref{Pontec} to counter-rotate 
the expectation value~\eqref{expneu} in the mass basis. 
In doing so, we  obtain
\begin{eqnarray}
\nonumber
A_1&=&i\,\mathcal{N}\hspace{0.1mm}G_F\, 
\delta^{3}\left(\bp_n-\bp_p-\bq-\bk\right)e^{-i\left(\omega_\bq^e\,+\,\omega_{\bk,1}\right)x^0_{out}}\cos\theta\,\bar{u}^s_{\bq,e}\,\gamma^{\mu}(\mathds{1}-\gamma^5)\\[2mm]
\nonumber
&&\times\int_{x_{in}^0}^{x_{out}^0}\,dx^0\,v^r_{\bk,1}\,e^{-i\left(\omega_{\bp_n}-\omega_{\bp_p}-\omega_{\bq}^e-\omega_{\bk,1}\right)x^0}\\[2mm]
&&\times\,V_{ud}\, \bar u_{\bp_p}^{\sigma_p}\hspace{0.3mm}\gamma_\mu (f\,-\,\gamma^5 g)\hspace{0.3mm} u_{\bp_n}^{\sigma_n}\,e^{-i(\omega_{\bp_p}x_{out}^0-\omega_{\bp_n}x^0_{in})}\,,
\label{amp1}
\end{eqnarray}
which becomes in the short-time limit
\begin{eqnarray}
\nonumber
A_1&\simeq&i\,\mathcal{N}\hspace{0.1mm}G_F\, 
\delta^{3}\left(\bp_n-\bp_p-\bq-\bk\right)\Delta t\,\cos\theta\,\bar{u}^s_{\bq,e}\,\gamma^{\mu}(\mathds{1}-\gamma^5)\,v^r_{\bk,1}\\[2mm]
&&\times\,V_{ud}\, \bar u_{\bp_p}^{\sigma_p}\hspace{0.3mm}\gamma_\mu (f\,-\,\gamma^5 g)\hspace{0.3mm} u_{\bp_n}^{\sigma_n}\,.
\label{Ap1apr}
\end{eqnarray}
This is predictably consistent with Eq.~\eqref{agSM} 
derived in the ultra-relativistic limit, where the effects of the mass difference can in fact be neglected. 

Similarly,  we infer the following expression for $A_2$
\begin{eqnarray}
\nonumber
A_2&=&i\,\mathcal{N}\hspace{0.1mm}G_F\, 
\delta^{3}\left(\bp_n-\bp_p-\bq-\bk\right)e^{-i\left(\omega_\bq^e\,+\,\omega_{\bk,2}\right)x^0_{out}}\sin\theta\,\bar{u}^s_{\bq,e}\,\gamma^{\mu}(\mathds{1}-\gamma^5)\\[2mm]
\nonumber
&&\times\int_{x_{in}^0}^{x_{out}^0}\,dx^0\,v^r_{\bk,2}\,e^{-i\left(\omega_{\bp_n}-\omega_{\bp_p}-\omega_{\bq}^e-\omega_{\bk,2}\right)x^0}\\[2mm]
&&\times\,V_{ud}\, \bar u_{\bp_p}^{\sigma_p}\hspace{0.3mm}\gamma_\mu (f\,-\,\gamma^5 g)\hspace{0.3mm} u_{\bp_n}^{\sigma_n}\,e^{-i(\omega_{\bp_p}x_{out}^0-\omega_{\bp_n}x^0_{in})}\,.
\label{amp2}
\end{eqnarray}
In the short-time approximation this yields
\begin{eqnarray}
\nonumber
A_2&\simeq&i\,\mathcal{N}\hspace{0.1mm}G_F\, 
\delta^{3}\left(\bp_n-\bp_p-\bq-\bk\right)\Delta t\,\sin\theta\,\bar{u}^s_{\bq,e}\,\gamma^{\mu}(\mathds{1}-\gamma^5)\,v^r_{\bk,2}\\[2mm]
&&\times\,V_{ud}\, \bar u_{\bp_p}^{\sigma_p}\hspace{0.3mm}\gamma_\mu (f\,-\,\gamma^5 g)\hspace{0.3mm} u_{\bp_n}^{\sigma_n}\,.
\label{Ap2apr}
\end{eqnarray}

To make a comparison with existing literature, 
it is convenient to consider  for a while the decay rates for the two
channels in Eq.~\eqref{masproc}, which are 
obtained by squaring the amplitudes $A_i$, $i=1,2$. Then, the approach is to compute the total
decay rate for the processes~\eqref{ndec} and~\eqref{ndecmu} through an incoherent sum of the contributions from the different mass eigenstates~\cite{GiuntiKim,suminc}, each averaged by $\cos^2\theta$ or $\sin^2\theta$, 
depending on whether one refers to the emission of an electron or muon
(anti-)neutrino\footnote{Notice that neutrinos can be produced incoherently if their mass differences are larger than the energy uncertainty in the production process.  This is quite plausible for heavy neutrinos ($m\gtrsim 100\,\mathrm{KeV}$ for typical energy of $10\,\mathrm{MeV}$).  
However, the present analysis is presumed to have a more general applicability than this specific case.}.
Following this recipe and resorting to Eqs.~\eqref{Ap1apr} and~\eqref{Ap2apr}, it is immediate to  
see that also in this framework
the decay probability for the process~\eqref{ndecmu}
is non-vanishing in the short-time limit, since it is given by the sum of two positive contributions. As already discussed for Pontecorvo states, this outcome is in contrast with what expected at tree level in the SM, 
which means that mass states are not eligible 
to be the fundamental representation for mixed neutrinos, 
if we want to preserve the internal consistency of the theory.

\section{Neutron $\beta$-decay with exact QFT flavor states}
\label{ExQFTs}
Let us now go through the above calculations by using 
exact QFT neutrino states. Toward this end, we
remind that these states are defined as eigenstates of flavor charges operators~\cite{BV95}. Indeed, by using the definition~\eqref{fqfts}, 
it is straightforward to check that~\cite{BV95}
\begin{eqnarray}
&:: Q_{\nu_e}(0)::|\lnu^{r}_{\;{\bf
k},e}\rangle\,=\,|\lnu^{r}_{\;{\bk},e}\rangle\,,\qquad :: Q_{\nu_\mu}(0)::|\lnu^{r}_{\;{\bf
k},\mu}\rangle\,=\,|\lnu^{r}_{\;{\bk},\mu}\rangle\,, \\[3mm]
&:: Q_{\nu_e}(0):: |\lnu^{r}_{\;{\bf
k},\mu}\rangle\,=\; :: Q_{\nu_\mu}(0)::|\lnu^{r}_{\;{\bk},e}\rangle\,=\,0\,,\qquad:: Q_{\nu_\ell}(0):: |0\rangle_{f} \,=\, 0\,,
\end{eqnarray}
where 
\be
Q_{\nu_{\ell}}(x^0)\,\equiv\,
\int d^{3}{\bx}\,\nu_{\ell}^{\dagger}(x)\hspace{0.3mm}\lnu_{\ell}(x) \,,\qquad \ell=e,\mu\,,
\label{fcnt}
\ee
are the flavor charge operators and $:: Q_{\nu_{\ell}} ::$ denotes the normal ordering respect  to the flavor vacuum. 
By contrast, it must be emphasized that 
Pontecorvo states (and, a fortiori, mass states) are not eigenstates of the flavor charges~\cite{Nucl}.

In this framework, the decay amplitude~\eqref{Aa} 
becomes
\be
A^{a)}\,=\,\langle\bar\nu^r_{\bk,e}, e^s_{\bq},\hspace{0.4mm} p^{\sigma_p}_{\bp_p} | 
\left[-i\int_{x^{0}_{in}}^{x^{0}_{out}} d^4x\, \mathscr{L}_{int}(x)
\right] |n^{\sigma_n}_{\bp_n}\rangle\,. 
\label{Aaexact}
\ee
The term involving the
expectation value of the neutrino field given
in Eq.~\eqref{ampP} is now replaced by
\begin{eqnarray}
\nonumber
\langle\bar\nu^r_{\bk,e} (x^0_{in})| \nu_e(x) | 0(x^0_{in})\rangle_f&\simeq& e^{-i\bk\cdot\bx}\left\{\cos^2\theta\, v_{\bk,1}^r\,e^{i\omega_{\bk,1}(x^0-x^0_{in})}\right.\\[2mm]
\nonumber
&&\,+\sin^2\theta\left[
\left(v^r_{\bk,1}\,|U_\bk|\,-\,\epsilon^r \,u^r_{-\bk,1}\,|V_\bk|
\right)|U_\bk|\,e^{i\omega_{\bk,2}(x^0-x^0_{in})}\right.\\[2mm]
&& \left.\left.+\left(v^r_{\bk,1}\,|V_{\bk}|\,+\,\epsilon^r\,u^r_{-\bk,1}\,|U_{\bk}|\right)|V_{\bk}|\,e^{-i\omega_{\bk,2}(x^0-x^0_{in})}\right]
\right\},
\label{exA}
\end{eqnarray}
which is defined 
consistently with the orthogonality of the Hilbert spaces
at different times, see footnote 3
(notice the presence
of the flavor vacuum and
the Bogoliubov coefficients $U_\bk$ and $V_\bk$).
Here we have used the field expansion~\eqref{flfiexp}
and the explicit form of the flavor annihilation/creation operators given in Eqs.~\eqref{boga2}-\eqref{bogc3}.  As above, normalization will be taken into account only at the end of calculations. 

By means of the identities~\eqref{prop3} and~\eqref{prop4}, 
Eq.~\eqref{exA} can be rewritten as
\begin{eqnarray}
\label{60}
\nonumber
\langle\bar\nu^r_{\bk,e} (x^0_{in})| \nu_e(x) | 0(x^0_{in})\rangle_f&\simeq&e^{-i\bk\cdot\bx}\left\{\cos^2\theta\, v_{\bk,1}^r\,e^{i\omega_{\bk,1}(x^0-x^0_{in})}\right.\\[2mm]
&& \left.+\sin^2\theta\left[v^r_{\bk,2}\,|U_\bk|\,e^{i\omega_{\bk,2}(x^0-x^0_{in})}\,+\,\epsilon^r\,u^r_{-\bk,2}\,|V_{\bk}|\,\,e^{-i\omega_{\bk,2}(x^0-x^0_{in})}
\right]
\right\}.
\end{eqnarray}
Combining Eqs.~\eqref{prot},~\eqref{elf} and~\eqref{60}, the decay amplitude~\eqref{Aaexact} takes the form
\begin{eqnarray}
\nonumber
A^{a)}&=&i\,\mathcal{N}\hspace{0.1mm}G_F\, 
\delta^{3}\left(\bp_n-\bp_p-\bq-\bk\right)e^{-i\omega_\bq^ex^0_{out}}\,\bar{u}^s_{\bq,e}\,\gamma^{\mu}(\mathds{1}-\gamma^5)\\[2mm]
\nonumber
&&\times\int_{x^0_{in}}^{x^0_{out}}\,dx^0\left\{\cos^2\theta\,v^r_{\bk,1}\,e^{-i\omega_{\bk,1}x^0_{in}}\,e^{-i\left(\omega_{\bp_n}-\omega_{\bp_p}-\omega_{\bq}^e-\omega_{\bk,1}\right)x^0}
\right.\\[2mm]
\nonumber
&&\,+\sin^2\theta\left[v^r_{\bk,2}\,|U_{\bk}|\,
\,e^{-i\omega_{\bk,2}x^0_{in}}\,e^{-i\left(\omega_{\bp_n}-\omega_{\bp_p}-\omega_{\bq}^e-\omega_{\bk,2}\right)x^0}\right.\\[2mm]
\nonumber
&&\left.\left.+\,\epsilon^r\,u^r_{-\bk,2}\,|V_{\bk}|\,
e^{i\omega_{\bk,2}x^0_{in}}\,e^{-i\left(\omega_{\bp_n}-\omega_{\bp_p}-\omega_{\bq}^e+\omega_{\bk,2}\right)x^0}
\right]\right\}\\[2mm]
&&\times\,V_{ud}\, \bar u_{\bp_p}^{\sigma_p}\hspace{0.3mm}\gamma_\mu (f\,-\,\gamma^5 g)\hspace{0.3mm} u_{\bp_n}^{\sigma_n}\,e^{-i(\omega_{\bp_p}x_{out}^0-\omega_{\bp_n}x^0_{in})}\,.
\label{AmpexactflA}
\end{eqnarray}
The structure of this amplitude is clearly different from the one in Eq.~\eqref{aa}, due to the presence of the Bogoliubov coefficients and the extra contribute of the anti-particle degrees of freedom (the term multiplying the $u$-neutrino mode)\footnote{The negative neutrino energy $\omega_{\bk,2}$ in the fourth line of Eq.~\eqref{AmpexactflA} should not be misleading, since it is associated to the ``hole'' contributions in the flavor vacuum condensate. Of course, this is a richness of the QFT treatment of mixing, which arises from the peculiar structure of flavor vacuum (see the discussion in the Introduction).}. Regardless of irrelevant phase factors, 
Eq.~\eqref{aa} is however recovered in the ultra-relativistic limit, where $|U_\bk|\rightarrow1$, $|V_\bk|\rightarrow0$ and exact flavor states reduce
to the standard Pontecorvo states.
We show below that the presence of such additional term in Eq.~\eqref{AmpexactflA}
is essential to restore the conservation
of lepton charge in the short-time limit and validate the use of exact QFT flavor states as correct neutrino representation.  

On the other hand, if we consider the decay channel in Eq.~\eqref{ndecmu}, the amplitude~\eqref{Amu} becomes
\be
A^{b)}\,=\,\langle\bar\nu^r_{\bk,\mu}, e^s_{\bq},\hspace{0.4mm} p^{\sigma_p}_{\bp_p} | 
\left[-i\int_{x^{0}_{in}}^{x^{0}_{out}}d^4x\, \mathscr{L}_{int}(x)
\right] |n^{\sigma_n}_{\bp_n}\rangle\,, 
\label{Abexact}
\ee
where the expectation value involving the neutrino field
is now given by
\begin{eqnarray}
\nonumber
\langle\bar\nu^r_{\bk,\mu} (x^0_{in})| \nu_e(x) | 0(x^0_{in})\rangle_f&\simeq& e^{-i\bk\cdot\bx}\,\sin\theta\cos\theta
\left[\left(v^r_{\bk,1}\,|U_\bk|\,-\,\epsilon^r \,u^r_{-\bk,1}\,|V_\bk|
\right)\,e^{i\omega_{\bk,2}(x^0-x^0_{in})}
\right.\\[2mm]
&&\left.-\,v^r_{\bk,1}\,|U_{\bk}|\,e^{i\omega_{\bk,1}(x^0-x^0_{in})}
\,+\,\epsilon^r\,u^r_{-\bk,1}\,|V_{\bk}|\,e^{-i\omega_{\bk,1}(x^0-x^0_{in})}
\right].
\end{eqnarray}
This can be simplified by use of the relation~\eqref{prop3} to give
\begin{eqnarray}
\nonumber
\langle\bar\nu^r_{\bk,\mu} (x^0_{in})| \nu_e(x) | 0(x^0_{in})\rangle_f&\simeq& e^{-i\bk\cdot\bx}\,\sin\theta\cos\theta
\left[v^r_{\bk,2}\,e^{i\omega_{\bk,2}(x^0-x^0_{in})}
\right.\\[2mm]
&&\left.-\,v^r_{\bk,1}\,|U_{\bk}|\,e^{i\omega_{\bk,1}(x^0-x^0_{in})}
\,+\,\epsilon^r\,u^r_{-\bk,1}\,|V_{\bk}|\,e^{-i\omega_{\bk,1}(x^0-x^0_{in})}
\right].
\end{eqnarray}
The amplitude~\eqref{Abexact} is then equal to
\begin{eqnarray}
\nonumber
A^{b)}&=&i\,\mathcal{N}\hspace{0.1mm}G_F\, \sin\theta\cos\theta\,
\delta^{3}\left(\bp_n-\bp_p-\bq-\bk\right)e^{-i\omega_\bq^ex^0_{out}}\,\bar{u}^s_{\bq,e}\,\gamma^{\mu}(\mathds{1}-\gamma^5)\\[2mm]
\nonumber
&&\times\int_{x^0_{in}}^{x^0_{out}}\,dx^0\,\left[v^r_{\bk,2}\,e^{-i\omega_{\bk,2}x^0_{in}}\,e^{-i\left(\omega_{\bp_n}-\omega_{\bp_p}-\omega_{\bq}^e-\omega_{\bk,2}\right)x^0}
\right.\\[2mm]
\nonumber
&&-\,v^r_{\bk,1}\,|U_{\bk}|\,e^{-i\omega_{\bk,1}x^0_{in}}\,e^{-i\left(\omega_{\bp_n}-\omega_{\bp_p}-\omega_{\bq}^e-\omega_{\bk,1}\right)x^0}
\\[2mm]
\nonumber
&&\left.+\,\epsilon^r\,u^r_{-\bk,1}\,|V_{\bk}|\,e^{i\omega_{\bk,1}x^0_{in}}\,e^{-i\left(\omega_{\bp_n}-\omega_{\bp_p}-\omega_{\bq}^e+\omega_{\bk,1}\right)x^0}\right]\\[2mm]
&&\times\,V_{ud}\, \bar u_{\bp_p}^{\sigma_p}\hspace{0.3mm}\gamma_\mu (f\,-\,\gamma^5 g)\hspace{0.3mm} u_{\bp_n}^{\sigma_n}\,e^{-i(\omega_{\bp_p}x_{out}^0-\omega_{\bp_n}x^0_{in})}\,.
\label{AmpexactflB}
\end{eqnarray}
Again, the ultra-relativistic limit gives back Eq.~\eqref{ab} up to
an irrelevant phase factor. 

\subsection{Short-Time Limit}
We now consider the short-time limit of the amplitudes~\eqref{AmpexactflA} and~\eqref{AmpexactflB}. Toward this end, 
we follow the same recipe as in Sec.~\ref{Aistl} 
and set $x^0_{in}=-\Delta t/2$ and $x^0_{out}=\Delta t/2$. 
By integrating over $x^0$, we get for $A^{a)}$
\begin{eqnarray}
\nonumber
A^{a)}&=&2i\,\mathcal{N}\hspace{0.1mm}G_F\, 
\delta^{3}\left(\bp_n-\bp_p-\bq-\bk\right)e^{-i\omega_\bq^e\,\Delta t/2}\,\bar{u}^s_{\bq,e}\,\gamma^{\mu}(\mathds{1}-\gamma^5)\\[2mm]
\nonumber
&&\times\left\{\cos^2\theta\,v^r_{\bk,1}\,e^{i\omega_{\bk,1}\Delta t/2}\
\frac{\sin\left[\left(\omega_{\bp_n}-\omega_{\bp_p}-\omega_{\bq}^e-\omega_{\bk,1}
\right)\Delta t/2
\right]}{\omega_{\bp_n}-\omega_{\bp_p}-\omega_{\bq}^e-\omega_{\bk,1}}\right.\\[2mm]\nonumber
&&+\,\sin^2\theta\left[v^r_{\bk,2}\,|U_\bk|\,e^{i\omega_{\bk,2}\Delta t/2}\
\frac{\sin\left[\left(\omega_{\bp_n}-\omega_{\bp_p}-\omega_{\bq}^e-\omega_{\bk,2}
\right)\Delta t/2
\right]}{\omega_{\bp_n}-\omega_{\bp_p}-\omega_{\bq}^e-\omega_{\bk,2}}
\right.\\[2mm]
\nonumber
&&\left.\left.+\,\epsilon^r\,u^r_{-\bk,2}\,|V_{\bk}|\,
e^{-i\omega_{\bk,2}\Delta t/2}\
\frac{\sin\left[\left(\omega_{\bp_n}-\omega_{\bp_p}-\omega_{\bq}^e+\omega_{\bk,2}
\right)\Delta t/2
\right]}{\omega_{\bp_n}-\omega_{\bp_p}-\omega_{\bq}^e+\omega_{\bk,2}}
\right]\right\}\\[2mm]
&&\times\,V_{ud}\, \bar u_{\bp_p}^{\sigma_p}\hspace{0.3mm}\gamma_\mu (f\,-\,\gamma^5 g)\hspace{0.3mm} u_{\bp_n}^{\sigma_n}\,e^{-i(\omega_{\bp_p}+\omega_{\bp_n})\Delta t/2}\,.
\end{eqnarray}

In the short-time approximation, this yields to the first order
\begin{eqnarray}
\nonumber
A^{a)}&\simeq&i\,\mathcal{N}\hspace{0.1mm}G_F\, 
\delta^{3}\left(\bp_n-\bp_p-\bq-\bk\right)\Delta t\,\bar{u}^s_{\bq,e}\,\gamma^{\mu}(\mathds{1}-\gamma^5)\,\left\{\cos^2\theta\,v^r_{\bk,1}\,+\,\sin^2\theta
\left[v^r_{\bk,2}\,|U_\bk|\,+\,\epsilon^r\,u^r_{-\bk,2}\,|V_{\bk}|
\right]
\right\}\\[2mm]
&&\times\,V_{ud}\, \bar u_{\bp_p}^{\sigma_p}\hspace{0.3mm}\gamma_\mu (f\,-\,\gamma^5 g)\hspace{0.3mm} u_{\bp_n}^{\sigma_n}\,,
\end{eqnarray}
which can be further manipulated by means of the identity~\eqref{prop6}
to give
\begin{eqnarray}
\nonumber
A^{a)}&\simeq&i\,\mathcal{N}\hspace{0.1mm}G_F\, 
\delta^{3}\left(\bp_n-\bp_p-\bq-\bk\right)\Delta t\,\bar{u}^s_{\bq,e}\,\gamma^{\mu}(\mathds{1}-\gamma^5)\,v^r_{\bk,1}\\[2mm]
&&\times\,V_{ud}\,\bar u_{\bp_p}^{\sigma_p}\hspace{0.3mm}\gamma_\mu (f\,-\,\gamma^5 g)\hspace{0.3mm} u_{\bp_n}^{\sigma_n}\,.
\label{finala}
\end{eqnarray}
As remarked above, this correctly matches the amplitude~\eqref{agSM}
calculated with Pontecorvo states in the
ultra-relativistic approximation. 

Let us now turn to the process~\eqref{ndecmu}. 
In the short-time limit, the amplitude~\eqref{AmpexactflB}
takes the form
\begin{eqnarray}
\nonumber
A^{b)}&\simeq&i\,\mathcal{N}\hspace{0.1mm}G_F\, \sin\theta\cos\theta\,
\delta^{3}\left(\bp_n-\bp_p-\bq-\bk\right)\Delta t\,\bar{u}^s_{\bq,e}\,\gamma^{\mu}(\mathds{1}-\gamma^5)\left[v^r_{\bk,2}\,-\,v^r_{\bk,1}\,|U_{\bk}|\,+\,\epsilon^r\,u^r_{-\bk,1}\,|V_{\bk}|\right]
\\[2mm]
&&\times\,V_{ud}\,\bar u_{\bp_p}^{\sigma_p}\hspace{0.3mm}\gamma_\mu (f\,-\,\gamma^5 g)\hspace{0.3mm} u_{\bp_n}^{\sigma_n}\,.
\label{finalb}
\end{eqnarray}

By comparison with Eq.~\eqref{absl}, we see that the main difference
is now the splitting of the mode with mass $m_1$ into two contributions multiplying the Bogoliubov coefficients $|U_\bk|$ and $|V_\bk|$, respectively. While the former survives in the ultra-relativistic limit (returning Eq.~\eqref{absl}), the latter has no counterpart in Pontecorvo framework, as it vanishes for $|V_\bk|\rightarrow0$.
However, it is exactly this term which allows to get
consistency with expectations from the SM at tree level.
Indeed, due to the identity~\eqref{prop3}, it is immediate to 
realize that the quantity in the square brackets 
is identically zero, which gives
\be
\label{Vanish}
A^{b)}=0\,.
\ee
This proves that the usage of 
the exact flavor states leads to the conservation
of lepton charge in the production vertex.
 
In order to better understand the above results, 
we observe that  
the amplitudes $A^{a)}$ and $A^{b)}$
in the short-time limit provide information 
on the decay processes in proximity
of the interaction vertex. We can think of associating a wave function $v^r_{\bk,\nu_e}$
to the emitted electron (anti-)neutrino in Eqs.~\eqref{App} and~\eqref{finala} (and, similarly, in Eqs.~\eqref{absl} and~\eqref{finalb}).
For the case of exact flavor states, Eq.~\eqref{finala}, 
one can identify $v^r_{\bk,\nu_e}=v^r_{\bk,1}$, with $v^{r\dagger}_{\bk,1}\,v^r_{\bk,1}=1$ being properly normalized (see Eq.~\eqref{ort}). 
By contrast,  Eq.~\eqref{App}
suggests that $v^r_{\bk\,\nu_e}=\cos^2\theta\,v^r_{\bk,1}\,+\,\sin^2\theta\,v^r_{\bk,2}$ for Pontecorvo states. However, 
by using Eqs.~\eqref{ort} and~\eqref{U}, 
simple calculations show that 
\begin{eqnarray}
\nonumber
v^{r\dagger}_{\bk\,\nu_e}\,v^r_{\bk\,\nu_e}&=&
(\cos^2\theta\,v^{r\dagger}_{\bk,1}\,+\,\sin^2\theta\,v^{r\dagger}_{\bk,2})\,(\cos^2\theta\,v^r_{\bk,1}\,+\,\sin^2\theta\,v^r_{\bk,2})\\[2mm]
\label{nnwf}
&=&\cos^4\theta\,+\,\sin^4\theta\,+\,2\sin^2\theta\,\cos^2\theta\,|U_\bk|\,,
\end{eqnarray}
from which we infer that the above wave function is not normalized, since $|U_{\bk}|<1$ for $m_1\neq m_2$. 
 
On the other hand, we notice that 
Eq.~\eqref{absl} includes the
wave function $v^r_{\bk,\Delta_{\nu_e}}=\sin\theta\cos\theta(v^r_{\bk,2}-v^r_{\bk,1})$. Not even this combination is 
normalized to unity, since 
\begin{eqnarray}
\nonumber
v^{r\dagger}_{\bk,\Delta_{\nu_e}}v^r_{\bk,\Delta_{\nu_e}}&=&\sin^2\theta\cos^2\theta(v^{r\dagger}_{\bk,2}-v^{r\dagger}_{\bk,1})\,(v^r_{\bk,2}-v^r_{\bk,1})\\[2mm]
&=&2\sin^2\theta\cos^2\theta\left(1-|U_{\bk}|\right)\,.
\end{eqnarray}
However, this is exactly the missing 
contribution to restore the normalization of the wave function~\eqref{nnwf}. Indeed, we have
\be
\label{norm1}
v^{r\dagger}_{\bk\,\nu_e}\,v^r_{\bk\,\nu_e}\,+\,v^{r\dagger}_{\bk,\Delta_{\nu_e}}v^r_{\bk,\Delta_{\nu_e}}\,=\,\cos^4\theta\,+\,\sin^4\theta\,+\,2\sin^2\theta\,\cos^2\theta\,|U_\bk|\,+\,2\sin^2\theta\cos^2\theta\left(1-|U_{\bk}|\right)\,=\,1.
\ee

A similar reasoning can be developed 
for mass states. In this case $v^r_{\bk\,\nu_e}=\cos\theta\,v^r_{\bk\,1}$ or $v^r_{\bk\,\nu_e}=\sin\theta\,v^r_{\bk\,2}$, depending on whether one considers Eq.~\eqref{Ap1apr} or~\eqref{Ap2apr}. 
Again, such wave functions are not normalized if taken separately. Nevertheless, the following condition holds
\be
\label{norm2}
\cos\theta^2\,v^{r\dagger}_{\bk\,1}v^{r}_{\bk\,1}\,+\,\sin\theta^2\,v^{r\dagger}_{\bk\,2}v^{r}_{\bk\,2}\,=\,1\,.
\ee

The above arguments clearly show that neither Pontecorvo nor mass states provide a self-consistent description of the neutron decay with neutrino mixing, as they are not exact eigenstates of the flavor charge. 
However, the reason why the sum over the two flavors (Eq.~\eqref{norm1}) or the two masses (Eq.~\eqref{norm2}) allows to recover the correct result can be understood by considering the time-dependent flavor charges~\eqref{fcnt} and the corresponding conserved charges for neutrinos with definite masses
\be
Q_{\nu_{i}}\,\equiv\,
\int d^{3}{\bx} \,
\nu_{i}^{\dagger}(x)\hspace{0.3mm}\lnu_{i}(x)\,\,,\qquad i=1,2\,.
\ee
In~\cite{BJV} it has been proved that 
\be
\sum_iQ_i\,=\,\sum_\ell Q_\ell(t)\,=\, Q\,,
\ee
where $Q$ represents the total charge of the system. 
The above equality can be interpreted as the conservation of the \emph{total lepton number}: when there is no mixing, this number is given by the sum of two separately conserved family lepton charges (left-hand side). On the other hand, in the presence of mixing the same conserved number
is obtained by adding time-dependent flavor charges, which are indeed associated to neutrino oscillations (central term).

In conclusion, we remark that our analysis 
is focused on the short-time limit of the $\beta$-decay amplitude. 
Clearly, a comprehensive study should also take
into account the opposite $\Delta t\rightarrow\infty$
limit. This has been investigated in~\cite{CYL},  
showing that the usage of exact QFT states
leads to predictions which are consistent with the SM model
for the case of relativistic neutrinos.

\section{Discussion and Conclusions}
\label{Disc}
In this work we have analyzed the flavor/mass dichotomy
for mixed neutrino fields. We have considered the neutron $\beta$-decay $n\,\rightarrow\, p \,+\,e^-\,+\,\bar\nu_e$ and the corresponding
flavor-violating channel $n\,\rightarrow\, p \,+\,e^-\,+\,\bar\nu_\mu$ as a test bench. For these processes, we have explicitly computed
the transition amplitude by using the scattering theory at tree level. Special focus has been devoted to the study of the short-time limit, i.e. the limit for very small distances from the interaction vertex. We have developed calculations by resorting to the following different representations of neutrino states: \emph{i}) 
Pontecorvo states, \emph{ii}) mass states and \emph{iii})
exact QFT flavor states, which are defined as eigenstates of the flavor charge. We have found that the usage of the latter
representation leads to results consistent with the 
conservation of lepton charge in the vertex, whereas
Pontecorvo and mass states fail. 
This provides a solid argument in favor of exact QFT states as neutrino fundamental representation. 
We stress that such a result has been achieved
based on the sole requirement of consistency with SM at tree level. In other terms, our formalism does not require any detector-dependent model for
the emission and absorption of neutrinos, nor 
a particular empirical description of
flavor oscillations~\cite{MatsasDet}.

In the above treatment, we have not calculated explicitly the 
neutron decay width. This should be taken into account
for a more direct comparison of our results with existing literature. Nevertheless, the fact that the amplitude $A_{n\rightarrow p+e^-+\bar\nu_\mu}$ computed with the
exact QFT flavor states identically vanishes
in the short-time limit is enough 
to infer that the decay width vanishes as well. 
By contrast, this does not happen when working with Pontecorvo states or, a fortiori, with mass states, for which the amplitude is in general nonzero. This outcome reveals unambiguously that 
both Pontecorvo and mass representations
are at odds with SM predictions.

Further aspects remain to be addressed in order to improve
the above analysis. 
Strictly speaking, we should have carried out calculations with three neutrino generations, including the possibility of $CP$ violation. 
Likewise, a description based on wave-packets 
would be more appropriate to take into account finite
spatial localization of neutrino production and detection.
However, we expect that such generalizations
do not affect the overall validity of our results. 
Furthermore, we have shown that 
the amplitudes computed with exact QFT and Pontecorvo states
are in good agreement in the ultra-relativistic limit.
Since observed neutrinos are essentially ultra-relativistic, 
it is quite difficult to find experimental evidences
that help to discern between the two representations at present. 
In spite of this, valuable hints could be provided by the PTOLEMY experiment~\cite{Betts}, which aims at detecting neutrinos of the cosmic background (CNB) through capture on tritium. Given that the temperature of the CNB is estimated around $T\approx 2\,\mathrm{K}$, it 
is reasonable to expect that these particles 
are mostly non-relativistic, thus offering a 
promising phenomenological window on the issue.

Finally, relevant advances in our understanding of neutrino mixing and oscillations in QFT might be achieved via the study of these phenomena 
in non-trivial spacetime. In this context, an emblematic 
research line is represented by the analysis of the 
accelerated proton decay with in Rindler metric~\cite{Ahluw,ProtBlas,ProtMat,GaetanoLucia}. 
Work along this direction is presently under active investigation 
and will be elaborated in future works.

\acknowledgements{The author is grateful to Massimo Blasone  for useful discussions. He also acknowledges the Spanish ``Ministerio de Universidades'' 
for the awarded Maria Zambrano fellowship and funding received
from the European Union - NextGenerationEU. He is grateful for 
participation in the COST Association Action CA18108  ``Quantum Gravity Phenomenology in the Multimessenger Approach'' and LISA Cosmology Working group. }

\end{document}